\documentclass[twocolumn,prb,showpacs]{revtex4}
\usepackage{psfrag}
\usepackage{graphicx}
\usepackage{amsmath}
\usepackage{amssymb}

\begin{document}
\title{Polarization-induced renormalization of molecular levels at metallic and semiconducting surfaces}
\author{J. M. Garcia-Lastra$^{1,2}$, C. Rostgaard$^1$, A. Rubio$^2$, and K. S. Thygesen$^1$}
\affiliation{$^1$Center for Atomic-scale Materials Design (CAMD)
  Department of Physics, Technical University of Denmark, DK - 2800 Kgs. Lyngby, Denmark\\
$^2$ETSF Scientific Development Centre,
Dpto. F\'{\i}sica de Materiales, Universidad del Pa\'{\i}s Vasco UPV/EHU, Centro
de F\'{\i}sica de Materiales CSIC-UPV/EHU and DIPC, Av. Tolosa 72, 20018 San
Sebasti\'{a}n, Spain}

\date{\today}

\begin{abstract}
On the basis of first-principles G$_0$W$_0$ calculations we study systematically 
  how the electronic levels of a benzene molecule are renormalized by substrate polarization when physisorbed on different
  metallic and semiconducting surfaces. The polarization-induced
  reduction of the energy gap between occupied and unoccupied
  molecular levels is found to scale with the substrate density of
  states at the Fermi level (for metals) and substrate band gap (for
  semiconductors). These conclusions are further supported by GW
  calculations on simple lattice models. By expressing the electron
  self-energy in terms of the substrate's joint density of states we
  relate the level shift to the surface electronic structure thus providing a microscopic explanation of the trends in the G$_0$W$_0$
  calculations. While image charge effects are not captured by
  semi-local and hybrid exchange-correlation functionals, we find that
  error cancellations lead to remarkably good agreement between the G$_0$W$_0$
  and Kohn-Sham energies for the occupied orbitals of the adsorbed molecule.
\end{abstract}

\pacs{85.65.+h,71.10.-w,73.20.-r,31.70.Dk} \maketitle 

\section{Introduction}
Solid-molecule interfaces are central to a number of important areas
of physics and chemistry including heterogeneous catalysis,
electrochemistry, molecular- and organic electronics, and scanning
tunneling spectroscopy\cite{rossmeisl,moth,nitzan03,stm_review}. Most
of our current understanding of level alignment at interfaces builds on
effective single-particle descriptions such as the Kohn-Sham scheme of density
functional theory (DFT)\cite{vasquez}. Within such theories the energy levels of a molecule close to a surface are determined by hybridization,
charge-transfer, and interface dipole fields -- all properties of the
static mean field potential defining the single-particle Hamiltonian.  On the other hand, from photo-emission
and electron transport measurements it is well known that the dynamic
polarizability of the molecule's local environment can have a large
influence on the level
positions\cite{johnson87,kubatkin,c60_ag,repp05,kahn03}.  Such
polarization effects, which are induced by changes in the charge state
of the molecule, are not captured by available single-particle
descriptions.

\begin{figure}[!h]
\begin{center}
\includegraphics[width=1.0\linewidth]{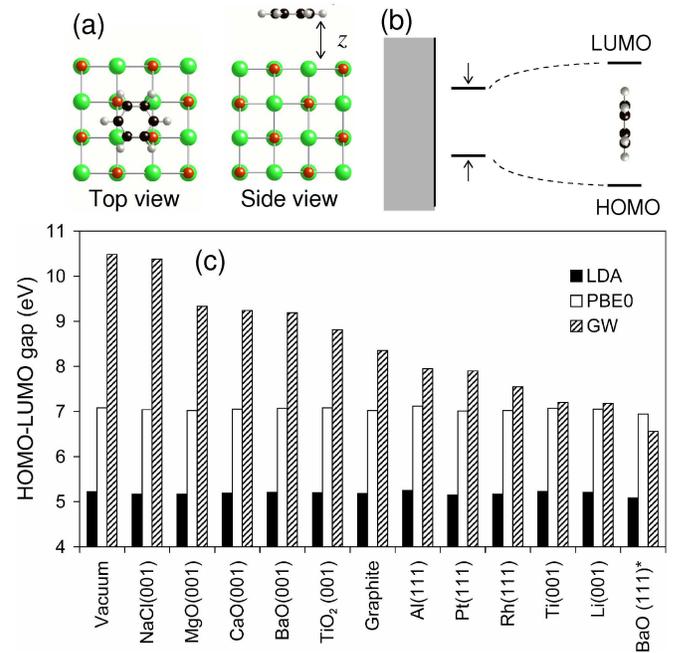}
\end{center}
\caption[system]{\label{fig1} (Color online) (a) Supercell used to represent benzene physisorbed on NaCl(001). (b) Reduction of a molecule's energy gap when it approaches a polarizable surface. (c) Calculated LDA, PBE0, and G$_0$W$_0$ HOMO-LUMO
  gap of a benzene molecule lying flat at $z=4.5$ \AA~above different
  surfaces. Note that BaO(111) is metallic due to surface states in the BaO band gap.}
\end{figure}

Many-body perturbation theory provides a systematic method to obtain
the true single-particle excitations [sometimes referred to as
addition/removal energies or quasiparticle (QP) energies] from the Green
function of the system. In the G$_0$W$_0$ approximation the electron
self-energy is written as a product of the (non-interacting) Green
function and a dynamically screened Coulomb interaction,
$\Sigma=iG_0W_0$\cite{gunnarsson,rmp}. It is instructive to compare this
to the bare exchange self-energy given by $\Sigma_x=iG_0V$, where $V$ is
the unscreened Coulomb interaction. It is well known that the Hartree-Fock (HF)
eigenvalues correspond to energy differences between the $N$-particle
groundstate and the \emph{unrelaxed} $N\pm 1$-particle
Slater determinants (Koopmans' theorem). The effect of replacing $V$ with the
screened and frequency dependent $W_0$ is two-fold: it introduces
correlations into the many-body eigenstates, and it includes the
response of the other electrons to the added electron/hole, i.e.
relaxation effects. For a molecule at a surface, the latter effect is particularly important as it
incorporates the attractive interaction between the added electron/hole and
its induced image charge, into the QP spectrum.

Recent experiments on molecular charge transport have renewed the interest for theoretical modeling of 
polarization-induced level renormalization.
First-principles G$_0$W$_0$ calculations for a benzene
molecule on graphite\cite{neaton} as well as CO on
NaCl/Ge(001)\cite{rinke} have demonstrated significant reductions of
the molecular energy gap due to image charge effects. Model GW
calculations have been used to elucidate the qualitative features
of the effect across different bonding
regimes\cite{gw_prl}. Classical electrostatic models of various
complexity have been developed to correct energy levels obtained
from single-particle
calculations.\cite{quek,mowbray,kaasbjerg,stadler_geskin}.

In this work, we present a systematic study of image charge-induced
renormalization at a range of different surfaces taking both a
classical and quantum many-body viewpoint.  We have performed DFT
calculations with local density approximation (LDA) and hybrid (PBE0)
exchange-correlation functionals as well as G$_0$W$_0$ calculations
for a benzene molecule weakly physisorbed on the metals Li, Al, Ti,
Rh, Pt, and the semiconductors/insulators TiO$_2$, BaO, MgO, CaO, and
NaCl. The results for the HOMO-LUMO gap of benzene are shown in Fig.~\ref{fig1}.
While LDA and PBE0 yields a substrate independent HOMO-LUMO gap, the
G$_0$W$_0$ gaps are reduced from the gas phase value by an amount
which depends on the polarizability of the surface. For all systems,
we find that the dependence of the QP gap on the distance to the
surface can be described by a classical image charge model.  However,
the model parameters are sensitive to the microscopic details of the
system and this limits the usefulness of the classical model in
pratice. By evaluating the G$_0$W$_0$ self-energy to second order we
obtain a simple analytic expression which relates the level shift to
the substrate's joint density of states weighted by Coulomb interaction matrix elements. The model suggests that the HOMO-LUMO gap
should scale with the substrate band gap (for semiconducting surfaces)
and density of states at the Fermi level (for metallic surfaces). This
trend is verified for the first-principles results and is further
supported by GW calculations for simple lattice models.  Finally, we
analyze the deviation between the DFT and G$_0$W$_0$ results in more
detail. We find that the occupied Kohn-Sham levels obtained with LDA (PBE0) are in very good agreement with the G$_0$W$_0$ results for benzene adsorbed on the metallic (semiconducting) surfaces, and we show that this is a result of significant error cancellation in the LDA/PBE0 approximations.

The paper is organized as follows. In Sec. \ref{sec.method} we outline
the methodology used for the first-principles and model GW calculations.
In Sec. \ref{sec.clas} we investigate to what extent the
first-principles G$_0$W$_0$ results can be explained by a classical
image charge model. In Sec. \ref{sec.micro} we derive a simple
analytical expression for the polarization-induced level shift and
show that it explains the main trends in both the first-principles as
well as the model calculations. At the end of the section we analyze
the description of occupied and unoccupied levels separately and
discuss the effect of error cancellations in the DFT
results. We conclude in Sec. \ref{sec.conclusion}

\section{Methods}\label{sec.method}

\subsection{\emph{Ab-initio} G$_0$W$_0$ calculations}
To model the solid-molecule interfaces we use a slab containing four
atomic layers of the substrate in the experimentally most stable
phase, and a benzene molecule lying flat above the
surface followed by 12\AA~of vacuum. The benzene molecule is not relaxed on the surface but is fixed in its gas phase structure at a distance $z$ from the surface. An example of a supercell is
shown in Fig. \ref{fig1}(a) for the case of benzene on
NaCl(001).  The number of atoms included in the supercell per atomic
layer is 9 for Al, Rh, Pt, Ti; 12 for Li and TiO$_2$; and 16 for NaCl,
MgO, CaO and BaO.  This corresponds to distances between periodically
repeated benzene molecules in the range 8.1 to 9.9~\AA. All DFT
calculations have been performed with the PWSCF code~\cite{baroni}
which uses norm-conserving pseudopotentials~\cite{fuchs}. For
exchange-correlation functionals we have used the local density
approximation~\cite{perdew_wang} as well as the PBE0 hybrid
functional\cite{pbe0,pbe0_2}. The Brillouin zone (BZ) was sampled on a 4x4x1
k-point mesh, and the wavefunctions were expanded with a cut-off energy of 40
Hartree.  

In the G$_0$W$_0$(LDA) method one obtains the QP energies from the linearized QP equation
\begin{equation}\label{eq.qpeq}
  \varepsilon_n^{\text{QP}} = \varepsilon_n^{\text{LDA}} + Z_n \langle
  \psi_n^{\text{LDA}}|\Sigma_{\text{GW}}(\varepsilon_n^{\text{LDA}})-v_{\text{xc}}|\psi_n^{\text{LDA}}
\rangle
\end{equation}
where $\psi_n^{\text{LDA}}$ and $\varepsilon_n^{\text{LDA}}$ are
LDA eigenstates and eigenvalues, and
\begin{equation}
  Z_n =\Big [1-\frac{\partial \langle
\psi_n^{\text{LDA}}|\Sigma_{\text{GW}}(\varepsilon)|\psi_n^{\text{LDA}}
\rangle}{\partial
\varepsilon}\Big |_{\varepsilon_n^{\text{LDA}}}\Big ]^{-1}.
\end{equation}
The self-energy, $\Sigma_{\text{GW}}$, is evaluated non-selfconsistently from the
single-particle Green function, i.e.  $\Sigma_{\text{GW}}=
iG_0W_0$, with $G_0(z)=(z-H^{\text{LDA}})^{-1}$. It is customary to use the random phase approximation for the screened interaction, i.e. $W_0=V(1-VP)^{-1}$ with $P=-iG_0 G_0$.

We have performed the G$_0$W$_0$ calculations with the Yambo code~\cite{yambo} using the LDA wavefunctions and eigenvalues from the PWSCF calculations as input.
The plasmon pole approximation has been applied with a frequency of 1 Hartree (the HOMO and
LUMO energies of benzene change by less than 0.05 eV when the plasmon
frequency is varied between 0.5 and 2.0 Hartrees).  In the calculation
of the self-energy we included a minimum of 200 empty states. We have
checked that calculations are converged with respect to slab
thickness, lateral supercell size, k-point mesh, all energy cut-offs,
and that we reproduce the results previously reported in Ref.
\onlinecite{neaton} for benzene on graphite at $z=3.25$ \AA.

\subsection{Model GW calculations}
\begin{figure}[!h]
\begin{center}
\includegraphics[width=0.8\linewidth]{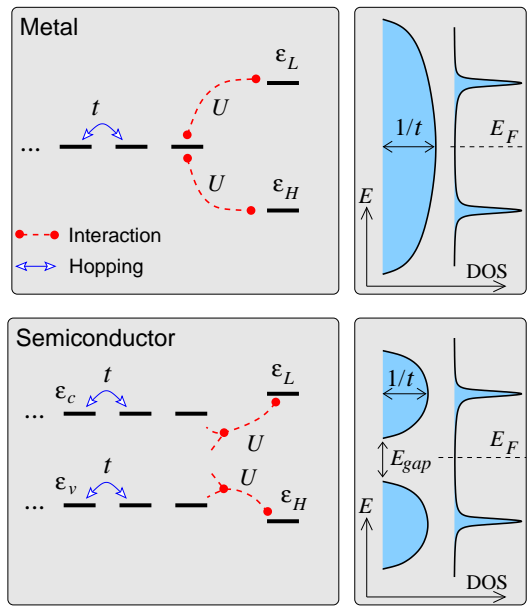}
\end{center}
\caption[system]{\label{fig2} (Color online) The lattice models
  representing a metal-molecule and semiconductor-molecule
  interface, respectively. We consider the weak coupling limit where no
  hybridization between the molecule and surface states occur. Thus the only interaction between the solid and molecule is via the non-local Coulomb interaction $U$.}
\end{figure}

In addition to the first-principles G$_0$W$_0$ calculations, we have
performed (self-consistent) GW calculations for two lattice models representing a
metal-molecule and semiconductor-molecule interface, respectively.
The model Hamiltonians contain three terms
\begin{equation}
\hat H=\hat
H_{\text{sol}}+\hat H_{\text{mol}}+\hat U,
\end{equation}
describing the solid (metal or semiconductor), the molecule, and their mutual interaction,
see Fig.~\ref{fig2}. A metallic substrate is modeled by a semi-infinite tight-binding
(TB) chain (we suppress the spin for notational simplicity),
\begin{equation}\label{eq.h_met}
\hat H_{\text{met}}=\sum_{i=-\infty}^0 t
(c^{\dagger}_{i}c_{i-1}+c^{\dagger}_{i-1}c_{i}).
\end{equation}
A semiconducting substrate is modeled by 
\begin{equation}\label{eq.h_met}
\hat H_{\text{sc}}=\sum_{\alpha=c,v}\sum_{i=-\infty}^0 \varepsilon_\alpha \hat n_{\alpha i}+t
(c^{\dagger}_{\alpha i}c_{\alpha i-1}+c^{\dagger}_{\alpha i-1}c_{\alpha i}),
\end{equation}
where $\alpha=c,v$ refers to conduction and valence bands, respectively.

The molecule is represented by its HOMO and LUMO levels, i.e. 
\begin{equation} 
\hat H_{\text{mol}}=\xi_{H}\hat
n_{H}+\xi_{L}\hat n_{L}
\end{equation}
where e.g. $\hat
n_{H}=c^{\dagger}_{H\uparrow}c_{H\uparrow}+c^{\dagger}_{H\downarrow}c_{H\downarrow}$,
is the number operator of the HOMO level. 

Finally, the interaction between the molecule and the terminal
site(s) of the substrate TB chain(s) is described by
\begin{displaymath} 
\hat U=\left \{ \begin{array}{ll}
U \hat n_0 \hat{N}_{\text{mol}} & \text{, for metals}\\
U \sum_\sigma(c^{\dagger}_{c0,\sigma}c_{v0,\sigma}+ c^{\dagger}_{v0,\sigma}c_{c0,\sigma})\hat{N}_{\text{mol}} & \text{, for semicond.}
\end{array}\right.
\end{displaymath}
where $\hat{N}_{\text{mol}}=\hat n_H+\hat n_L$ is the number operator of
the molecule. Note that since polarization of a semiconductor occurs via transitions between valence and conduction bands, 
only the interaction terms of the form given above contribute to the image charge effect (this will become clear in Sec. \ref{sec.2nd}).

We set $E_F=0$ corresponding to a half filled band for the metal. We
choose $\xi_H$ and $\xi_L$ so that the molecule contains exactly two
electrons ($E_F$ in the middle of the HOMO-LUMO gap).  We consider the
limit of zero hybridization between the solid and molecule so that
interaction between the solid and molecule occurs only via the non-local $\hat U$. The model neglects
interactions within the TB chain and between the molecule and interior
TB sites ($i<0$). These approximations are, however, not expected to
influence the image charge physics described by the model in any
qualitative way.

We obtain the Green function of the molecule from 
\begin{equation}
G(z)=1/(z-H_{\text{mol}}-\Sigma_{\text{GW}}[G](\varepsilon))
\end{equation}
where the Hartree potential due to $\hat U$ has been absorbed in $H_{\text{mol}}$. 
The GW self-energy is calculated fully
self-consistently using a recently developed GW scheme for quantum transport\cite{gw_prb}.
The renormalized molecular QP levels are obtained as peaks in the spectral function $A_{\nu}(\varepsilon)=-(1/\pi) \text{Im}G^r_{\nu  \nu}(\varepsilon)$.

\section{Classical Theory}\label{sec.clas}
In this section we investigate to what extent the G$_0$W$_0$ results of Fig. \ref{fig1} can be described by a classical image-charge model.

The electrostatic energy of a point charge, $q$, located in
vaccum at position $(0,0,z)$ above a polarizable medium filling the
half-space $z<z_0$, is given by (in atomic units)
\begin{equation}\label{eq.clas}
V=\frac{qq'}{4(z-z_0)}.
\end{equation}
The size of the image charge is $q'=q(1-\epsilon)/(1+\epsilon)$, where
$\epsilon$ is the relative dielectric constant of the
medium\cite{rohlfing03}. In 1973 Lang and Kohn showed that the energy
of a classical point charge above a quantum jellium surface follows
Eq. (\ref{eq.clas}) with $q'=-q$ (corresponding to $\epsilon=\infty$
as expected for a perfect metal), with the image plane, $z_0$, lying 0.5-0.9
\AA~outside the surface depending on the electron
density\cite{lang_kohn}. More recently, \emph{ab-initio} G$_0$W$_0$
calculations have found the same asymptotic form for the potential
felt by an electron outside a metallic
surface\cite{rohlfing03,eguiluz92,white98}. From this it seems reasonable to conclude that the
assymptotic position of the electronic levels of a molecule outside a surface
would also follow the image potential of Eq. (\ref{eq.clas}). This is,
however, only true for the unoccupied levels whereas the occupied
levels experience a shift in the opposite direction, i.e. the
shift is upward in energy as the molecule approaches the surface. This is because the occupied levels represent the negative of the energy cost of removing an electron from the molecule. Similarly it has been found that 
the image potential leads to band gap narrowing at
semiconductor-metal interfaces.\cite{inkson,charlesworth}.

\begin{figure}[!h]
\begin{center}
\includegraphics[width=1.0\linewidth]{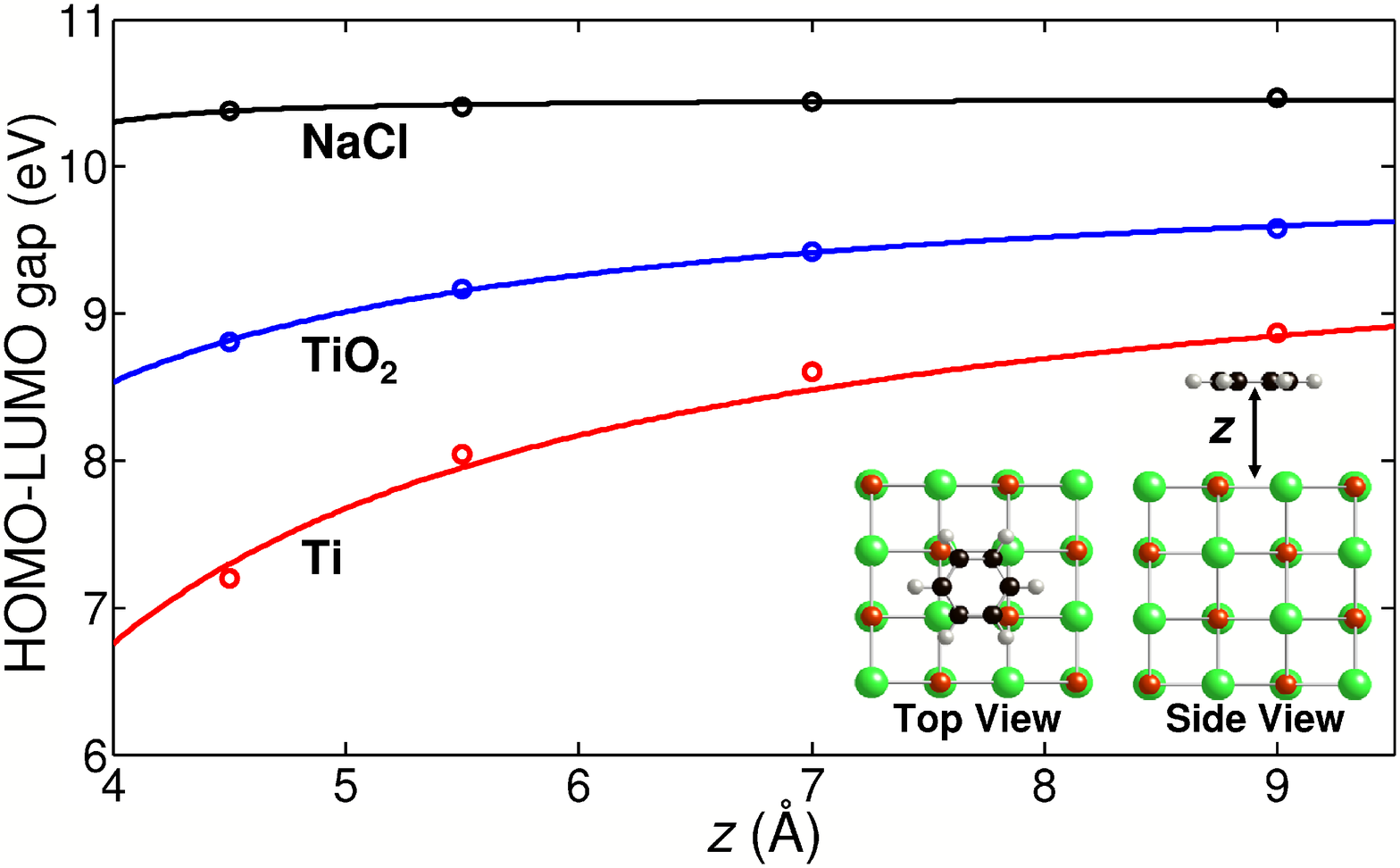}
\end{center}
\caption[system]{\label{fig3} (Color online) Calculated G$_0$W$_0$ energy gap of benzene on NaCl, TiO$_2$, and Ti surfaces (circles) as a function of the distance to the surface, and the best fit to the classical model Eq. (\ref{eq.clas}) (full lines).}
\end{figure}

To test whether the gap reductions obtained in the G$_0$W$_0$ calculations can be described by the classical
image charge model we have fitted Eq. (\ref{eq.clas}) to the calculated HOMO-LUMO
gap for $z=4.5, 5.5, 7.0,9.0, \infty$ \AA. In
Fig. \ref{fig2} we show the result of the fit for three systems (the
fit is equally good for the other systems). The best-fit values for
the effective image plane $z_0$ and the dielectric constant
$\epsilon_{\text{model}}$ are
given in table \ref{table1}. 

As can be seen $\epsilon_{\text{model}}$
is generally smaller than the experimental optical dielectric constant of the bulk,
$\epsilon_{\infty}^{\text{exp,bulk}}$. This is expected since the latter
gives the long-range response of the bulk while
$\epsilon_{\text{model}}$ probes the local response at the surface. Part of the discrepancy between $\epsilon_{\infty}^{\text{exp}}$ and $\epsilon_{\text{model}}$ is clearly due to geometric effects. By taking the surface geometry into account, as done in Ref. \onlinecite{kaasbjerg}, better estimates of $\epsilon_{\text{model}}$ can be produced from $\epsilon_{\infty}^{\text{exp}}$. 
On the other hand, electronic effects due to the local atomic structure of the surface cannot be captured by a classical model. For example, the BaO(111) surface is metallic due to surface states, and thus
$\epsilon_{\text{model}}\approx \infty$ while
$\epsilon_{\infty}^{\text{exp}}=3.83$. Similarly, impurities, defects, and
surface roughness are expected to influence the local dielectric
properties of the surface. 

According to the classical image charge model all the molecular levels
should experience the same shift (the sign of the shift being
different for occupied and unoccupied levels). However, we have found
that the best-fit values for $z_0$ and $\epsilon_{\text{model}}$
obtained by fitting the HOMO and LUMO levels speparately, are in
general different -- most notably for the metallic surfaces. This observation, which is discussed in more detail
in Sec. \ref{sec.orbital}, shows that the shape of molecular orbital
also influences the size of the polarization-induced shift.

\begin{table}[!h]
\caption{Position of the effective image plane, $z_0$, and dielectric constant, $\epsilon_{\text{model}}$, obtained by fitting the $z$-dependence of the HOMO-LUMO gap to Eq. (\ref{eq.clas}). Last row shows the experimental optical dielectric constant of the bulk. The two values for the non-isotropic TiO$_2$ refers to longitudinal and transverse polarization directions.}
\begin{center}
\renewcommand{\arraystretch}{1.2}
\begin{tabular}{l|c|c|c} \hline\hline
                         & $z_0$(\AA)  & $\epsilon_{\text{model}}$  & $\epsilon^{\text{exp,bulk}}_{\infty}$  \\
\hline
% --------------------------------------------------------------------------------
NaCl(001)               & 1.70 & 1.15 & 2.30   \\
MgO(001)		& 1.20 & 2.63 & 2.95   \\
CaO(001)		& 2.69 & 1.56 & 3.30 \\
BaO(001)		& 2.74 & 1.77 & 3.83 \\
TiO$_2$(001)		& 1.79 & 2.76 & 8.43/6.84 \\
Al(111)			& 0.55 & $\infty$ & $\infty$ \\
Pt(111)			& 0.60 & $\infty$ & $\infty$ \\
Rh(111)			& 1.28 & $\infty$ & $\infty$ \\
Ti(001)			& 1.66 & $\infty$ & $\infty$ \\
Li(001)			& 1.72 & $\infty$ & $\infty$ \\
BaO(111)		& 2.01 & $\infty$ & 3.83 \\
% --------------------------------------------------------------------------------
\hline\hline
\end{tabular}
\flushleft
Experimental data taken from Ref. \onlinecite{oxmat}\\
\label{table1}
\end{center}
\end{table}

\section{Microscopic theory}\label{sec.micro}
In this section we first consider the GW self-energy for a molecule interacting with a surface to second-order in the electron-electron interaction. This leads to a simple microscopic model for image charge renormalization which relates the shift of molecular levels to the electronic structure of the surface, and explains general trends of the first-principles and model GW calculations. In the last
section we consider the HOMO and LUMO levels separately and explain how error cancellations in semi-local exchange-correlation functionals can explain the surprisingly good agreement found between LDA eigenvalues and GW QP energies for the occupied levels of benzene on metallic surfaces.

\subsection{Second-order expansion}\label{sec.2nd}
\begin{figure}[!h]
\begin{center}
\includegraphics[width=1.0\linewidth]{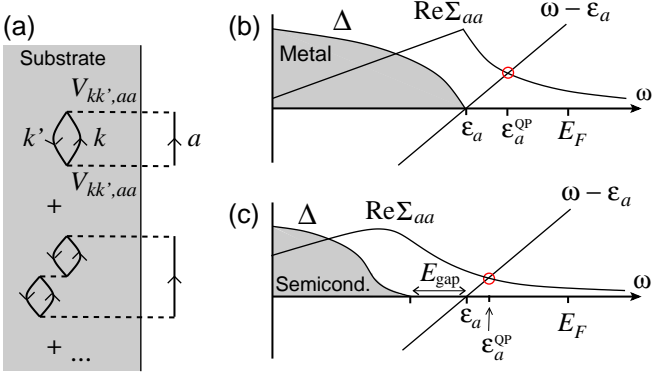}
\end{center}
\caption[system]{\label{fig4} (a) Feynman diagrams representing
  dynamic polarization of the substrate induced by an electron
  propagating in the molecule. (b) and (c): Generic shapes of the
  imaginary and real parts of the self-energy of Eq.
  (\ref{eq.sigmafinal}) for an occupied molecular level $|a\rangle$ interacting with a metallic and semi-conducting substrate
  assuming $V_{kk',aa}$ to be energy independent.}
\end{figure}

In quantum many-body theory, the effect of substrate polarization on the energy levels of a molecule enters the Green function via a self-energy operator. In general, the G$_0$W$_0$ self-energy can be written symbolically as  
\begin{equation}
\Sigma=\sum_{n=1}\Sigma^{(n)}=\sum_{n=1}iG_0V(PV)^{n-1},
\end{equation}
where $G_0$ is the Green function of the non-interacting (Kohn-Sham)
Hamiltonian, and $P=-iG_0G_0$ is the polarization bubble. The first-order
term, $\Sigma^{(1)}$, is simply the static exchange potential while
the remaining terms account for correlations and dynamic
screening. In the following we consider the second-order term,
$\Sigma^{(2)}=iG_0VP V$ explicitly. This corresponds to approximating the reponse of the substrate by its non-interacting response, $P$.

For sufficiently large surface-molecule separations ($z\gtrsim 3.5$\AA) we can neglect hybridization effects, and the non-interacting eigenstates
of the combined system can be taken as the eigenstates of the isolated
molecule and surface. We denote these eigenstates by $\{\psi_a\}$ ('$a$' for adsorbate) and
$\{\psi_k\}$, respectively.  To see how a given electronic
level, $\varepsilon_a$, is renormalized by polarization processes in
the substrate we consider the (time-ordered) matrix element
$\Sigma_{a a}^{(2)}(\omega)=\langle
\psi_a|\Sigma^{(2)}(\omega)|\psi_a\rangle$, given by
\begin{equation}\label{eq.sigma}
\Sigma_{a a}^{(2)}=\sum_{k}^{\text{occ}} \sum_{k'}^{\text{empty}}\!\!\!\int iG_{0,a a}(\omega') V_{aa,k k'\
}P_{kk'}(\omega'-\omega) V_{k' k,aa}  \text{d}\omega'.
\end{equation}
The Feynman diagram corresponding to $\Sigma_{a a}^{(2)}$ is shown in Fig. \ref{fig4}(a). The polarization and Coulomb matrices are given by
\begin{eqnarray}
P_{kk'}(\omega)&=&\frac{1}{\omega-\omega_{kk'}+i\eta}-\frac{1}{\omega+\omega_{kk'}-i\eta}\\
V_{k k',aa} &=& \int\!\!\! \int \frac{\psi_k^*(\bold r) \psi_{k'}(\bold r)|\psi_a(\bold r')|^2}{|\bold r-\bold r'|}\text{d}
\bold r \text{d}\bold r'
\end{eqnarray}
where $\eta$ is a positive infinitesimal and
$\omega_{kk'}=\varepsilon_{k'}-\varepsilon_k\geq 0$. Using that $G_{0,a
a}(\omega)=1/(\omega-\varepsilon_a
+\text{sgn}(\varepsilon_a-E_F) i\eta)$~\cite{gunnarsson}, Eq. (\ref{eq.sigma}) reduces to
\begin{equation}\label{eq.sigma2}
\Sigma_{a a}^{(2)}(\omega)=\frac{1}{\pi}\int  \frac{\Delta(\omega')}{\omega-\omega' +\text{sgn}(E_F-\varepsilon_a) i\eta} d\omega'
\end{equation}
where we have defined the interaction strength,
\begin{equation}\label{eq.delta}
\Delta=\pi \sum_{k}^{\text{occ}} \sum_{k'}^{\text{empty}} |V_{kk',aa}|^2 \delta(\omega_{kk'}-\text{sgn}(\varepsilon_a-E_F)(\omega-\varepsilon_a)).
\end{equation}
Note that $\Delta$ is simply the joint density of states (JDOS) of the substrate, shifted by $\varepsilon_a$, and weighted by the Coulomb matrix elements. The physically relevant retarded self-energy is readily obtained from Eq. (\ref{eq.sigma2})
\begin{eqnarray}\label{eq.sigmafinal}
\Sigma_{a a}^{(2),r}=\frac{\mathcal P}{\pi}\int  \frac{\Delta(\omega')}{\omega'-\omega} d\omega'-i \Delta(\omega).
\end{eqnarray}
where $\mathcal P$ denotes the Cauchy principal value. Now, the renormalized QP
energy can be obtained from the equation (neglecting off-diagonal terms)
\begin{equation}
\varepsilon_a^{\text{QP}}-\varepsilon_a-\langle \psi_a|\Sigma^{(2),r}(\varepsilon_a^{\text{QP}})|\psi_a\rangle=0
\end{equation}
A graphical solution to the QP equation is illustrated in Fig.
\ref{fig4}(b,c) for the case of an occupied molecular level
$\varepsilon_a<E_F$ interacting with a metal or semiconductor
surface, respectively.

From Eq. (\ref{eq.delta}) it follows that the image charge effect does not 
broaden the molecular level because 
$\text{Im}\Sigma^{(2)}(\varepsilon_a)=0$. We also note that the
level shift is independent of the absolute value $|\varepsilon_a-E_F|$, and that the effect of changing the sign
of $\varepsilon_a-E_F$ is to change the sign of the level shift.
These properties are all in line with the classical theory. 

In the limit where $V_{kk',aa}$ varies little with $k$ and $k'$, $\Delta$
is simply proportional to the shifted JDOS [the "generic" cases
illustrated in Figs. \ref{fig4}(b,c)]. In this case the level shift is simply determined by the form of the JDOS.
For a metal, the JDOS raises linearly at $\omega=0$ with a slope given by
the metal's DOS at $E_F$. This suggests that the level shift should
increase with the substrate DOS at the Fermi level. For a semiconductor, the JDOS raises smoothly at
$\omega=E_{\text{gap}}$, suggesting that the level shift should decrease
with $E_{\text{gap}}$. In the following section we investigate these relations for the model and first-principles calculations. We mention that the second order approximation discussed above may not always provide a good description of the full GW self-energy. However, as we will show in the next section, it explains qualitatively the trends in G$_0$W$_0$ calculations.

\subsection{Dependence of level shift on surface electronic structure}
\begin{figure}[!h]
\begin{center}
\includegraphics[width=1.0\linewidth]{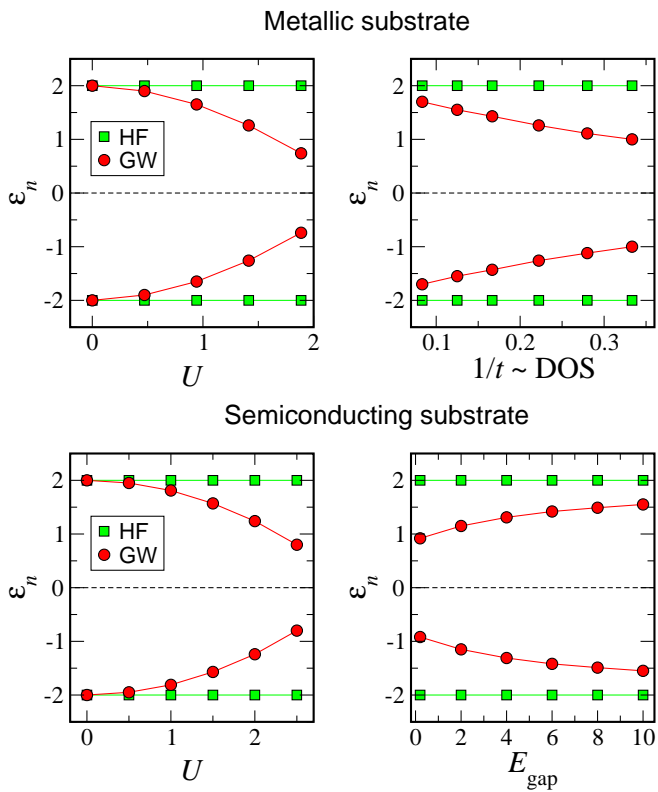}
\end{center}
\caption[system]{\label{fig5} HOMO and LUMO positions obtained from the simple lattice models for a metallic substrate (upper panel) and semiconducting substrate (lower panel). In all plots we vary one parameter while keeping the remaining parameters fixed\cite{parameters}.} 
\end{figure}

In Fig. \ref{fig5} we show the HOMO and LUMO levels of the lattice
models calculated with the HF and GW approximations. In all plots we vary one parameter
of the model while keeping the remaining parameters
fixed\cite{parameters}. 

\begin{figure}[!h]
\begin{center}
\includegraphics[width=1.0\linewidth]{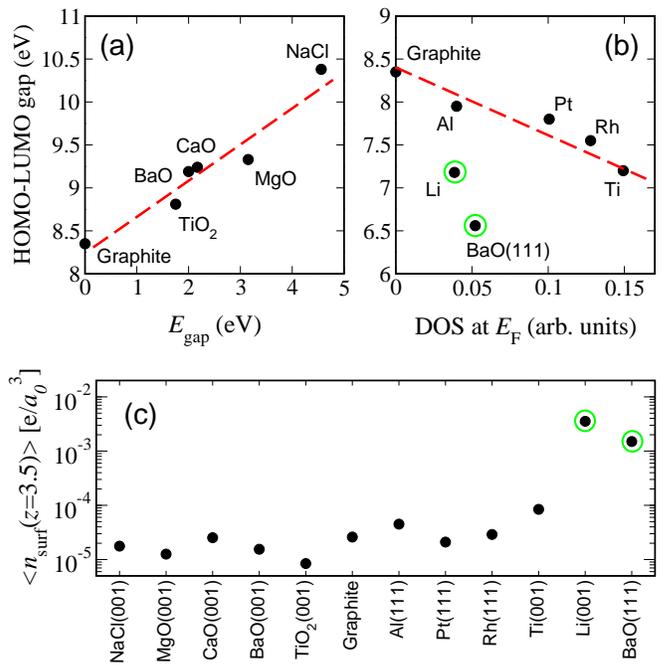}
\end{center}
\caption[system]{\label{fig6} Calculated
  HOMO-LUMO gap of benzene at $z=4.5$ \AA (same numbers as in Fig.
  \ref{fig1}) plotted as function of the LDA substrate band gap for
  the semiconductors (a), and the total DOS per volume evaluated at the
  Fermi level for the metals (b). Dashed lines have been added to guide the eye. (c) Average electron density in a plane lying $z=3.5$ \AA above the clean surfaces.}
\end{figure}

The upper panels refer to a metallic substrate and show the dependence
on the levels on the interaction strength $U$ and the intra-chain
hopping parameter $t$. Note that the latter is inversely proportional
to the projected density of states (DOS) of the terminal site
evaluated at $E_F$. The lower panels refer to a semiconducting
substrate and show the dependence of the levels on $U$ and the
substrate gap, $E_{\text{gap}}$. The HF eigenvalues are clearly
independent of the non-local interaction between the molecule and
substrate. This can be understood from Koopmans' theorem which states
that the HF eigenvalues do not include the electronic relaxations of
the substrate induced by the extra electron/hole in the molecule. In
contrast the GW levels vary in the way predicted by the simple model
discussed in the previous section: The polarization-induced reduction
of the HOMO-LUMO gap is stronger for larger $U$ as well as for larger
substrate DOS at $E_F$ for the metals and smaller substrate band gap
for the semiconductors. A more detailed discussion of level renormalization based on the
lattice model for metallic substrates, including the case of strong
metal-molecule hybridization, can be found in Ref. \onlinecite{gw_prl}.
 
In Fig. \ref{fig6}(a,b) we plot the G$_0$W$_0$ gaps from Fig. \ref{fig1}
versus the LDA band gap and DOS at $E_F$ for the semiconducting and
metallic substrates, respectively. For the semiconductors the
reduction of the HOMO-LUMO gap clearly correlates with
$E_{\text{gap}}$. This indicates that the interaction strength, i.e. the
matrix elements $V_{kk',aa}$ of Eq. (\ref{eq.delta}), do not differ too much from one
surface to another. For the metals, the HOMO-LUMO gap seems to scale
with the metal's DOS at $E_F$. However, we note that Li(001) and BaO(111)
deviate from the general trend followed by the other metals.
This can be explained by the larger extend of the metallic
wavefunctions of these systems into the vacuum region, which in turn
leads to larger $V_{kk',aa}$ matrix elements. Indeed, Fig. \ref{fig6}(c) shows the average electron density evaluated in a plane lying $z=3.5$ \AA above the surface in the absence of the benzene molecule. The density outside the Li(001) and BaO(111) surfaces is significantly larger than for the other surfaces which on the other hand have quite similar densities.

\subsection{DFT eigenvalues and error cancellation}\label{sec.orbital}
In Fig. \ref{fig3} we plot the energies of the HOMO and LUMO levels of
benzene at $z=4.5$ \AA. For each surface, we have shifted the LDA,
PBE0, and G$_0$W$_0$ levels by the same amount so that the LDA HOMO is
aligned with the HOMO in the gas phase. We note that the effect of
substrate polarization is very similar for the G$_0$W$_0$ HOMO
and LUMO levels which are shifted up and down, respectively, by almost
the same amount.  This is indeed expected from the classical image
charge model.  Significant deviations from this trend are, however,
seen for Li(001) and BaO(111). We ascribe this to the more extended
nature of the metallic states on these surfaces which reduce the
validity of the point charge approximation and can introduce
differences between the $V_{kk',HH}$ and $V_{kk',LL}$ matrix elements.

Overall, the LDA and PBE0 eigenvalues for the HOMO are in better
agreement with the G$_0$W$_0$ QP energies than is the case for the
LUMO. Moreover there is a general trend that the LDA eigenvalues come
closer to the G$_0$W$_0$ energies as we move from the insulating to
the metallic surfaces. In fact, the LDA HOMO level is almost on top of
the G$_0$W$_0$ level on the metallic surfaces.  This trend is clearly
a result of sigificant error cancellation in the LDA. Indeed, it is
well known that semilocal exchange-correlation functionals
overestimate (underestimate) occupied (empty) molecular levels due to
self-interaction effects. At the metallic surfaces this error is
compensated by the missing image charge correction. PBE0 gives better
estimates for the free molecule where it opens up the LDA HOMO-LUMO
gap due to partial removal of self-interaction errors. In this case,
the cancellation between the missing image charge effect and the
remaining self-interaction error results in very good agreement
between PBE0 and G$_0$W$_0$ for the HOMO level on the semiconducting
surfaces.

The cancellation between self-interaction errors and missing polarization effects will always be present in hybrid- and semilocal approximations. However, the relative size of the
two contributions will in general depend on the shape of the molecule, its
orientation with respect to the surface, the molecule-surface distance, and the type of substrate.

\begin{figure}[!h]
\begin{center}
\includegraphics[width=1.0\linewidth]{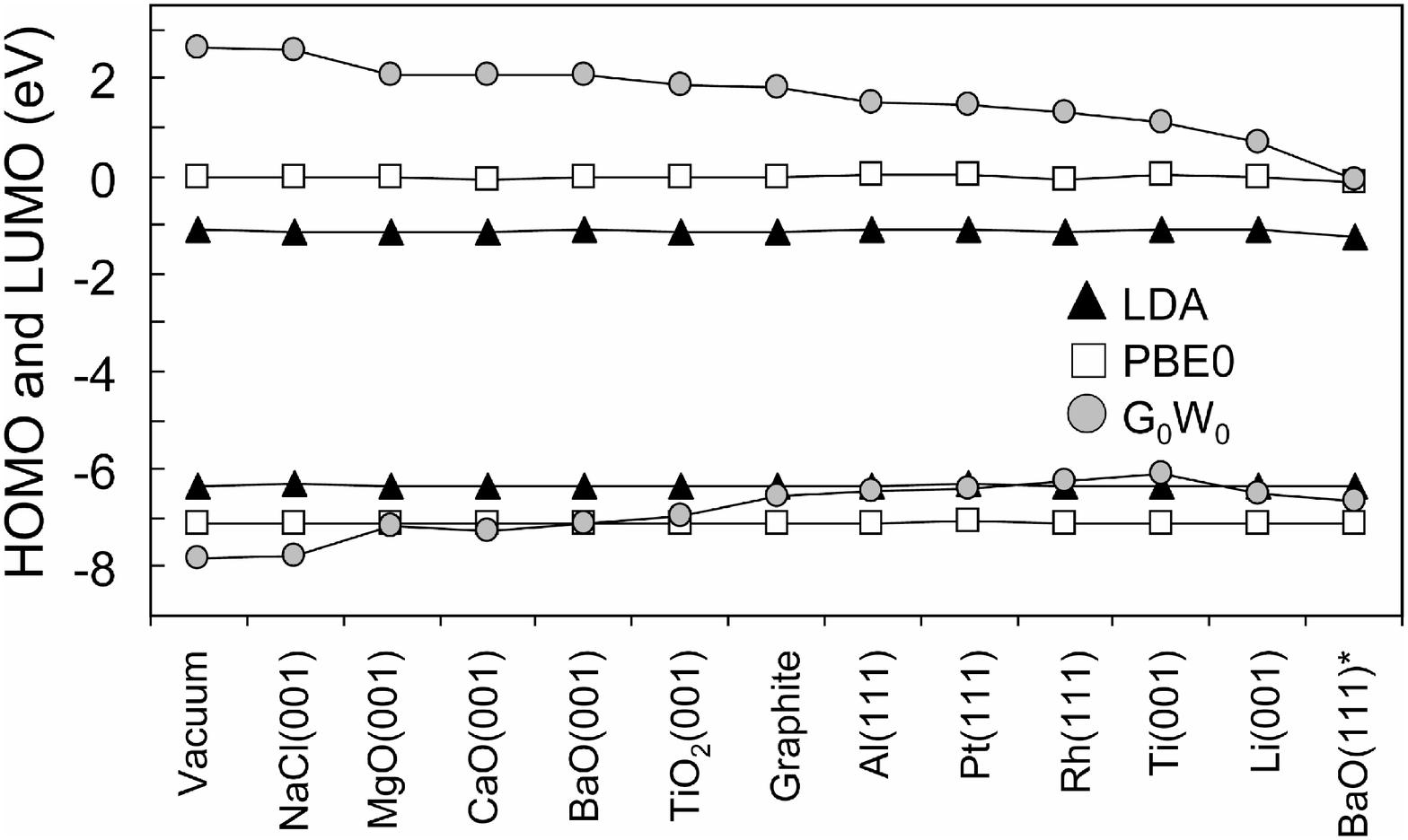}
\end{center}
\caption[system]{\label{fig3} LDA, PBE0, and G$_0$W$_0$ energies for the HOMO and LUMO levels of benzene at $z=4.5$~\AA~above the surfaces. The very good agreement between LDA and G$_0$W$_0$ energies for the HOMO level at the metallic surfaces is due to error cancellation in the LDA approximation.}
\end{figure}

\section{Conclusions}\label{sec.conclusion}
We have presented G$_0$W$_0$ calculations for a benzene molecule
physisorbed on different metallic and semiconducting surfaces. Upon
physisorption the molecule's HOMO-LUMO gap is reduced from its gas
phase value due to dynamic polarization of the substrate. It was shown
that a classical image charge model captures the qualitative features
of the effect while the magnitude of the level shift is sensitive to
the detailed atomic structure of the surface. In particular the
presence of metallic mid-gap state at the surface of a semiconductor
can have a large influence on the local response of the surface. Both
local and hybrid exchange-correlation potentials fail to account for
the polarization effects yielding Kohn-Sham eigenvalues of physisorbed
benzene which are independent of the substrate. Nevertheless we found
that a cancellation between self-interaction errors and missing image
charge effects in the LDA leads to a very good agreement between LDA
and G$_0$W$_0$ energies for the occupied states of benzene on metallic
surfaces. Similar conclusions were reached for the PBE0 energies on
semiconducting substrates. Finally, we have derived a simple
second-order approximation to the GW self-energy which expresses the
polarization-induced shift of a molecular level in terms of the
substrate's joint density of states weighted by Coulomb interaction
matrix elements. This model was used to explain general trends in the
first-principles results, namely the scaling of the benzene's
HOMO-LUMO gap with the substrate density of states at $E_F$ (for
metals) and the substrate band gap (for semiconductors).
 
Our results clearly demonstrates the importance of non-local
correlations for the electronic levels at solid-molecule interfaces.
We expect this to have important implications for the theoretical
modelling of electron transport in organic and single-molecule devices.

\section{Acknowledgements}
KST and CR acknowledge support from the Danish Center for
Scientific Computing. The Center for Atomic-scale
Materials Design (CAMD) is sponsored by the
Lundbeck Foundation. AR and JMGL acknowledge funding by the Spanish MEC (FIS2007-65702-C02-01),
"Grupos Consolidados UPV/EHU del Gobierno Vasco" (IT-319-07), e-I3 ETSF
project (Contract Number 211956) and "Red Espa\~nola de Supercomputaci\'on".

%%%%%%% References

\end{document}